\documentstyle[aas2pp4,flushrt,eqsecnum]{article}

\def\Rquart{$R^{1/4}$}

\begin{document}

\title{USING FUNDAMENTAL PLANE DISTANCES to ESTIMATE\\
the TOTAL BINDING MASS in ABELL 2626\altaffilmark{1}}

\author{Joseph J. Mohr}
\affil{Departments of Physics and Astronomy, University of Michigan, Ann Arbor, MI 48109}
\authoremail{jjmohr@umich.edu}
\author{Gary Wegner}
\affil{Department of Physics and Astronomy, Dartmouth College, Hanover, NH  03755}
\authoremail{wegner@kayz.dartmouth.edu}

\altaffiltext{1} {Observations reported here were obtained at the 
MDM Observatory, a facility jointly operated by the
University of Michigan, Dartmouth College, and the Massachusetts Institute
of Technology.}

\bigskip
\centerline{Accepted for publication in {\it The Astronomical Journal}}

\begin{abstract}
We use fundamental plane (FP) distance estimates to the components
of the double cluster A2626 ($cz\sim17,500$~km/s)
to constrain cluster kinematics and estimate total binding mass.
We employ deep $R$ band CCD photometry, multi--object
spectroscopy, and software designed to account for seeing effects to
measure the FP parameters $R_e$, $\sigma$, and $\left<\mu_e\right>$
for 24 known early type and S0 cluster members. 
The FP coefficients from this sample
($\alpha=1.30\pm0.36$ and $\beta=0.31\pm0.06$)
are consistent with others reported in the literature.

We examine the $Mg_b$ equivalent width distributions within
both subclusters and find them to be indistinguishable.
Lacking evidence for stellar population differences, we interpret the
FP zeropoint offset between the two subclusters as a 
measure of the distance difference.
We find $\log(D_B/D_A)=-0.037\pm0.046$,
where $D_{cl}$ is the distance to subcluster $cl$.  This measurement
is consistent with the subclusters being at the same distance, and it
rules out the Hubble flow hypothesis
(distances proportional to velocity) with 99\% confidence;
analysis of the subcluster galaxy magnitude distributions rules out
Hubble flow at 93\% confidence. Both results favor a kinematic model where the
subclusters are bound and infalling.

We estimate the total cluster binding mass by
modelling the subcluster merger as radial infall.
The projected separation, the line of sight velocity difference and the line
of sight separation constrain the cluster mass;  the {\it minimum} possible
total binding mass is 1.65 times higher than the sum of the standard
virial masses, a difference statistically significant at the
$\sim3\sigma$ level.  We discuss explanations for the inconsistency
including (1) biases in the standard virial mass estimator,
(2) biases in our radial infall mass estimate,
and (3) mass beyond the virialized cluster region;
if the standard virial mass is significantly in
error, the cluster has an unusually high mass--to--light ratio ($\sim1000h$).
Because observational signatures of departures from radial
infall are absent, we explore the implications
of mass beyond the virialized, core regions.

\end{abstract}


\section{Introduction}

Galaxy cluster masses are clearly of cosmological significance, and
examples abound: (1) the differences in cluster evolution as a
function of mass provide
constraints on the power spectrum (\cite{lac93}, 1994),
(2) comparisons of
the amounts of luminous and non--luminous matter in clusters provide
clues to the nature of dark matter and the efficiency of galaxy formation
(\cite{dav95}), and (3) the cluster mass--to--light ratio and 
baryon fraction constrain the cosmological
density parameter $\Omega_0$ (e.g. \cite{ram89,whi93,moh96a}).

Cluster mass studies tend to focus on the central
regions where virial equilibrium is a more accurate approximation,
and the weak lensing signals are easiest to detect.
Here we apply a method to measure the mass beyond the virialized region
in the double cluster A2626. 
We utilize recent improvements in the understanding of the
fundamental plane (FP) to constrain the total cluster
binding mass by revisiting the subcluster radial
infall model (\cite{bee82}). 
Specifically, measuring FP
distances (with an accuracy of 15\% to 25\%) to a reasonably large sample of
cluster galaxies can yield a very accurate
(sub)cluster distance (\cite{luc88,lyn88}).  For two
bound subclusters, the line of sight separation, the line of sight velocity
difference, and the projected separation provide enough information to
estimate the total cluster binding mass in cases where a radial infall model
is appropriate.  

Although A2626 is rather distant, it is well suited for 
this analysis because simply demonstrating that the two subclusters are
bound and infalling has interesting implications.
A straightforward application of the radial infall model,
using the (rest frame) line of sight velocity difference and the virial mass
estimates within each subcluster, rules out bound and infalling merger models
(\cite{moh96b}; hereafter MGW96);  the virial masses are not large enough to
produce the apparent velocity difference.
Thus, FP distances used to determine whether or not the subclusters in A2626 are
bound and infalling serve as an independent
test of the virial estimators, or alternatively,
as a means of measuring the mass beyond the virialized region of the cluster.

Possible pitfalls to this approach include (1) distance biasing FP 
zeropoint (or shape) differences between subclusters,
(2) deviations from radial infall, and (3) significant
interactions between the virialized regions of the two subclusters. 
FP variation with cluster environment is an area of active research (e.g.
\cite{guz92,wor96,jor96}; hereafter JFK96); current
indications are that distance biases are smaller than 5\% for ellipticals
within clusters and are related to observable variations in
the stellar populations within cluster galaxies. 
As discussed below, subcluster pairs suffering from pitfall numbers
2 and 3 can be identified through corresponding observational signatures.
We argue that the radial infall model and FP distances provide a promising
method for determining cluster masses on scales larger than
the virialized region in low redshift clusters; if coupled with estimates of
cluster light at larger radii, these mass measures should suffice to determine
whether mass to light ratios vary significantly between the virialized core
and the surrounding infall region.  For higher redshift cluster
pairs, our approach could be used to quantify the level of environmentally
induced fundamental plane zeropoint variations.

We describe the double cluster A2626, the FP observations, and the reductions
in $\S2$. In Section 3, we derive FP coefficients, discuss constraints on
the stellar populations in each subcluster, and derive constraints
on their relative distances. Section 4 contains
a discussion of the radial infall model applied to A2626.  Conclusions
are summarized in Section 5.  Throughout the paper we use $H_0=100h$~km/s/Mpc.

\section{Galaxy Sample and Data}

Below we describe Abell 2626 (\cite{abe58}), summarize previous
observational results detailed elsewhere (MGW96),
present the new observations, and describe the reductions
and analysis in detail.

\subsection{Abell 2626}

\begin{figure}
\plotfiddle{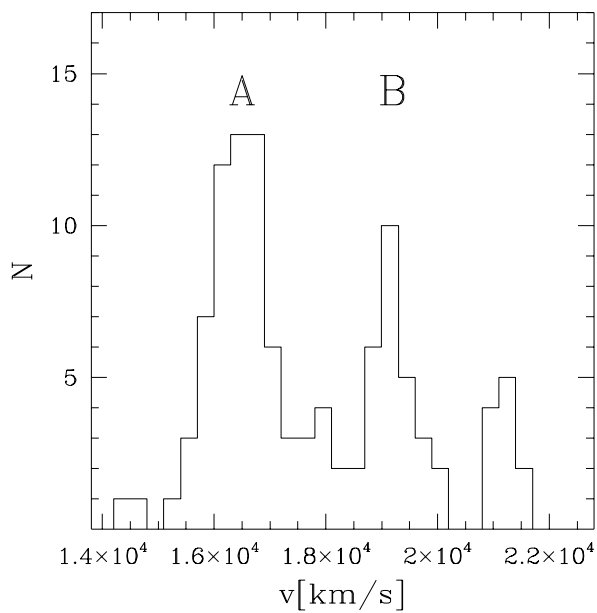}{5in}{0}{90}{90}{-180}{-370}
\caption{The galaxy velocity distribution around Abell~2626.
The histogram contains 108 galaxy velocities from the redshift survey of
MGW96.  Group A is centered
at $v=16,500$~km/s, and group B is centered at $v=19,150$~km/s. 
The high fraction of galaxies with emission lines 
in the group of 11 centered at $v\sim21,500$~km/s
(73\% compared to 39\%/33\% for groups A/B)
indicates they are part of a low density
background structure.\label{figvel}}
\end{figure}

A redshift survey with the Decaspec (\cite{fab90}) mounted on the MDM 
Hiltner 2.4~m telescope
in Fall 1993 and 1994 revealed that Abell 2626
($\left<cz\right>\sim17,500$~km/s, richness class 0) is composed of at least
two systems with mean line of sight velocities which differ by
$\sim2,500$~km/s (MGW96).  As is clear from
Fig. \ref{figvel} and Table \ref{grouptab},
there are two
main components with cluster--like dispersions:  group A centered
at a velocity of $\left<cz\right>=16,533\pm141$~km/s and group B centered
at a velocity of $\left<cz\right>=19,164\pm138$~km/s (90\% statistical
confidence limits). 
As discussed in MGW96, the
11 galaxies centered at $\left<cz\right>=21,173\pm119$~km/s are most likely
part of a background, low density structure because (1) the dispersion
is low ($\sigma=200^{+119}_{-52}$~km/s) and (2) a large fraction of these
galaxies are gas rich--- 73\% have emission lines compared to 39\% (33\%)
for group A (B).  If the velocity of the low density structure corresponds
to Hubble flow then its distance from groups A and B is from 20$h^{-1}$~Mpc
to 50$h^{-1}$Mpc.

\placefigure{figvel}

The group A and B masses, 
velocity dispersions,
central densities and X--ray properties all differ. 
As noted in 
Table \ref{grouptab},
the virial
mass of subgroup A is roughly twice that of B, and the line of sight
velocity dispersions are $\sigma_A=658^{+111}_{-81}$~km/s and
$\sigma_B=415^{+117}_{-72}$~km/s (90\% statistical confidence limits).  The
absence of observed X--ray emission from group B can be used to place
a limit on the central
gas density in group B (MGW96).  In particular, the {\it Einstein} 
imaging proportional counter (IPC) image of the region reaches
roughly 50 times fainter than the peak in the X--ray emission from group A.
The group A central gas density is $\sim3\times10^{-3}$~cm$^{-3}$,
so the central gas density in group B must be $\le5\times10^{-4}$~cm$^{-3}$.
Although the contrast in the projected galaxy densities of groups A and B
is difficult to quantify, the smoothed distribution (Figure 3 in MGW96)
indicates that the central density in group B is $\sim7$ times less than
the central density in group A. 
Thus, the ratio of gas to galaxy density may be the same in both groups.

\placetable{grouptab}

As discussed below, if groups A and B are
on a radial infall trajectory, then the minimum implied gravitational
mass is more than the sum of their virial masses.  However, it is also
possible that the velocity difference is just due to Hubble flow.  The
magnitude distributions of the confirmed members of both groups indicate
that the merger hypothesis is favored over Hubble flow at 93\%
confidence (MGW96).  To further investigate this issue we measure
fundamental plane distances to each cluster and determine the line of
sight distance difference directly.  To do so we first use $R$ band CCD
photometry to identify a sample of early type galaxies with
redshift confirmed membership.

\subsection{Photometry}

On November 21, 1995, we used a thinned, Tektronix 1024$^2$ CCD
mounted on the MDM Hiltner 2.4~m telescope
to image galaxies in A2626 and J8 (\cite{jac82}). 
The 24$\mu$ pixels are 0.275~$''$ on a side.  The seeing varied
during the night between 0.85~$''$ and 1.05~$''$. 
Because the night was non--photometric, we 
determine the photometric zeropoint of each $R$ band image
externally.  For the A2626 galaxies we use photometric images taken
previously (MGW96) on the MDM~1.3~m to zeropoint our images.  Specifically,
we use aperture photometry of isolated stars in the 2.4~m and 1.3~m
frames to determine the 2.4~m zeropoint.  The 1.3~m images are reduced to
the Kron--Cousins system using Landolt (1992) standards.
For J8 we rely primarily on published photoelectric
galaxy aperture photometry (\cite{col93}) to determine the
R band zeropoint but also use stellar aperture photometry from a photometric
1.3~m image for one galaxy (\cite{sag96}).  In addition to our primary imaging
run, we acquired additional Hiltner 2.4m images through the generosity of
Paul Schechter; the reductions for these data were similar.

The $R$ band galactic extinction is 0.062~mag toward A2626
(NASA/IPAC Extragalactic Database, \cite{sav79}) and
0.185~mag toward J8 (\cite{sag96}).  We approximate the $k$
correction as $k_R=z_g$ where $z_g$ is the
galaxy redshift (\cite{fre94}).  We account for
the peculiar velocity component of the
line of sight velocity when applying the cosmological dimming correction;
specifically, two of the factors of $(1+z)$ in the 
cosmological dimming are due to relativistic effects and
two are due to the change in the geometry of the universe between emission
and detection.  Writing $z_g=z_d+z_p$, where $z_d$ and $z_p$ are the Hubble
flow and peculiar velocity components of the redshift, the cosmological
dimming is written $C=+5\log{\{1+z_d\}}+5\log{\{1+z_g\}}$;
in terms of the observed and peculiar redshift, the correction is
$C=+10log{\{(1+z_g)\sqrt{1-z_p/(1+z_g)}\}}$.  Thus, $C$ is model
dependent.  As an example, if groups A and B are at the same distance
and the peculiar velocity correction were incorrectly taken
to be $z_p=0$, the fundamental plane distances would be biased so that
group B would appear closer than group A by $\sim1$\%.

We bias subtract, flat field (using twilight flats), and
clean the images using standard IRAF tasks.
For each galaxy, we calculate sky subtracted radial profiles
and uncertainties.  Uncertainties have a Poisson 
component added in quadrature with an assumed 1\% flat
fielding uncertainty.  We fit these profiles to summed \Rquart laws
and exponential disks,
\begin{eqnarray}
I_{b}(R) = I_{b0} e^{-7.67\left[(R/R_b)^{1/4}-1\right]} \\
I_{d}(R) = I_{d0}e^{-R/R_d}\nonumber
\end{eqnarray}
where $I_{b0}$ ($R_{b}$) and $I_{d0}$ ($R_{d}$) are the characteristic surface
brightnesses (scale lengths).
Our fitting procedure accounts for the effects of PSF smoothing and 
pixel extent as described in \cite{sag93b}.

More specifically, we determine the parameters of the \Rquart and
disk components
by minimizing the $\chi^2$ difference between the
seeing convolved model and the observed profile.  The expected surface
brightness $I(R,dR)$ in the bin of width $2dR$ and radius R is
\begin{equation}
{I(R,dR) = {\left[F(R+dR) - F(R-dR)\right] \over 4\pi R dR}}
\end{equation}
where $F(R)$ is the seeing convolved integral flux within radius $R$.
The integral flux follows from the bulge, disk and PSF parameters.
Specifically,
\begin{equation}
{F(R) = R \int_0^\infty dk J_1(kR) \hat p(k) \left( \hat I_{b}(k)
+ \hat I_{d}(k)\right)} 
\end{equation}
where $J_1$ is a Bessel function, and $\hat p(k)$, $\hat I_{b}(k)$,
and $\hat I_{d}(k)$ are the Fourier transforms of the PSF,
the bulge, and the exponential disk.  Following \cite{sag93b},
we use a PSF of the form
\begin{equation}
{\hat p(k) = e^{-(kb)^\gamma}}
\end{equation}
and find it to be a good fit to stars on the 2.4~m images
with $1.4\le\gamma\le1.6$.
We use the analytic approximation to the Fourier transform of the
\Rquart law (\cite{sag93b}), and the exact form of the
transform of the disk component (e.g. \cite{bra86}).

The fitting procedure has three steps:  
\begin{itemize}
\item Fit the PSF parameters $b$ and $\gamma$ using radial profiles
of $\sim4$ stars within each 2.4~m image;
best fit parameters minimize the $\chi^2$ difference
between the observed and model profile.

\item Feed initial guesses for the
bulge and disk parameters into a simplex minimization routine
(\cite{pre88}) which minimizes the $\chi^2$ difference
between the seeing convolved theoretical and observed profiles. 
At each iteration the
full Fourier integral is solved for each point in the radial profile;
with an HP--735 the minimization takes several minutes.  In addition
to the uncertainties in the observed profile, a theoretical
uncertainty is introduced to account for imperfections in the smoothing
corrections (e.g. the exact form of the PSF).  Specifically,
the uncertainty for each profile point has a component which
is proportional to the fractional effect of the smoothing on the model
profile.  This has the positive effect of making the fit results less sensitive
to the inclusion/exclusion of the heavily smoothed inner points in the profile
and to the details of the smoothing operation.
We include a theoretical uncertainty which is 10\% of
the fractional effect of the smoothing on the model profile.
This is consistent with a philosophy that we can correct for
seeing effects at the 90\% level.

\item We fit a bulge only model to the profile and compare the minimum
$\chi^2$ for this model to the bulge plus disk $\chi^2$.  For
galaxies where $\chi^2$ is only minimally improved by the inclusion of the
disk component we use the bulge only fit.
Once the fitting is complete, we use the best fit parameters to calculate
$R_e$-- the half luminosity radius, $\mu_e$-- the surface brightness at
$R_e$, and $\left<\mu_e\right>$-- the mean surface brightness within $R_e$.
\end{itemize}

We test our fitting routines by (1) fitting artificial galaxy images,
(2) comparing results from multiple images of the same galaxy, and
(3) comparing our results to those of Saglia et al. (1997)
in the cluster J8.  We create artificial images by adding a galaxy
described by an \Rquart\ profile to a flat background of $10^3$ cts/pixel.
We then smooth the artificial image and introduce Poisson noise.  By using
delta functions instead of \Rquart\ profiles we produce stars which are
then used to measure the PSF parameters $b$ and $\gamma$.  The process
of fitting artificial galaxy images
tests the accuracy of our fitting in that case where the true galaxy
profile is well described by an \Rquart\ profile.  We find that the parameters
$R_e$ and $\left<\mu_e\right>$ are constrained to an accuracy
of $\Delta \log(R_e)\sim0.02$
and $\Delta\left<\mu_e\right>\sim0.07$ for $R_e$ as small as 50\% the
FWHM of the PSF.  As expected, the accuracy of the combination
$\log(R_e)-0.33*\left<\mu_e\right>$
(which appears in the equation describing the fundamental plane) is
constrained to $\sim0.01$ because the errors in $R_e$ and $\left<\mu_e\right>$
are correlated.

We also compare results from subarcsecond seeing images of
seven A2626 galaxies to results from poorer quality images.
The mean offset between the parameters in the good and poor seeing images 
of the same galaxies are consistent with zero, and the scatter is
small.  Specifically, $\Delta\left<\mu_e\right>=-0.0262\pm0.0311$ with
an RMS of 0.0823, $\Delta \log(R_e)=-0.0073\pm0.0074$ with an
RMS of 0.0196, and $\Delta\left(\log(R_e)-0.33*\left<\mu_e\right>\right)=
0.0013\pm0.0037$ with an RMS of 0.0097.

We observed five galaxies in the EFAR cluster J8 (\cite{weg96})
for which we have photometric zeropoints.
These observations provide the strongest test of our methods because the
data, reductions and analyses differ significantly.
There are no significant offsets between our values and EFAR
values (\cite{sag96}); specifically, 
$\Delta\left<\mu_e\right>=-0.0766\pm0.1223$ with
an RMS of 0.2735, $\Delta \log(R_e)=-0.0257\pm0.0326$ with an
RMS of 0.0729, and $\Delta\left(\log(R_e)-0.33*\left<\mu_e\right>\right)=
-0.0006\pm0.0112$ with an RMS of 0.0249.
These three tests of our photometric methods indicate that
our procedure is accurate and makes only a small contribution to the observed
scatter in the fundamental plane (see below).

\placetable{galtab}

\subsection{Spectroscopy}

On November 22 and 23, 1995, we used the Decaspec 
(\cite{fab90}) plus MkIII spectrograph
mounted on the MDM Hiltner 2.4~m telescope to obtain spectra of four galaxy fields in
Abell 2626, one field in J8 (\cite{weg96}),
and multiple spectra of five HD stellar velocity
standards.  The Decaspec fibers subtend 2.3$''$ in the focal plane.
We used a grism blazed at 5,700~\AA\ with 600~l/mm,
yielding a typical PSF of 5.3~\AA\ at 2.18~\AA\ per pixel and coverage
from 4,300~\AA\ to 6,500\AA.
Four or five thirty minute exposures were obtained for each of the
A2626 fields; shorter exposures were obtained for the brighter J8 field.
We extracted the spectra, using arc lamp exposures taken on either side of
each object exposure to dispersion correct. We then combined the spectra from the
four to five exposures, excluding cosmic rays through sigma clipping of
large, positive deviations.
With the Decaspec there are four sky spectra for each
object spectrum.  We combined the sky spectra for each object,
scaled the sky spectrum using the flux in the 5577~\AA\ sky line, and
then removed the sky contribution from the object spectrum.
The final spectra range in signal to noise (S/N) from S/N=20 to 75 per
pixel at 5,300~\AA.

We obtained spectra of more galaxies, stellar velocity and Lick
linestrength standards (\cite{wor94b}) using the Modspec mounted on the
MDM Hiltner 2.4m telescope
in Nov and Dec '96.  We used a 1,200~l/mm grating, the 200~mm camera,
a thinned CCD with 24~$\mu$m pixels, and a 1.7$''$ wide slit; standard
longslit reductions led to spectra with 2~\AA\ resolution, 1~\AA/pixel
sampling, and coverage from 4,800\AA\ to 5,800\AA.  Galaxy exposure times ranged
from 20~min to 2.5~hrs, and the spectra have S/N$\ge30$ per pixel.
The higher resolution of this setup allowed us to accurately measure galaxy
dispersions as low as 100~km/s.

For each run, galaxy dispersions and velocities are extracted
using the stellar templates from that run
and the cross correlation peak fitting available in fxcor (IRAF.RV); we use the
approximate wavelength region 4,800--5,800\AA\ which contains the strong Mg
features.  The correlation peak
width is transformed to an intrinsic galaxy dispersion
using translation tables created for each stellar template; the translation
tables are produced by cross correlating the stellar templates
against Gaussian broadened versions of themselves.
The final galaxy dispersion is the mean of the 
dispersions from each stellar template in the run.
Finally, an aperture correction
\begin{equation}
\label{dispcor}
{\log{\left(\sigma_{cor}\over\sigma_{obs}\right)} =
0.038\log{\left({r_{ap}\over r_{ap}^{n}}{\theta_e^{n} \over\theta_e}\right)}}
\end{equation}
is applied. This aperture correction accounts for the falling
dispersion profile in early type galaxies; we normalize to an aperture
of radius $r_{ap}^{n}=1.7''$ and a galaxy with $\theta_e^{n}=20''$. 
For the longslit
spectra we use $r_{ap}=1.025\sqrt{xy/\pi}$, where $x$ and $y$ are the width
and length of the rectangular aperture (\cite{jor95,bag96}). 

We estimate the uncertainties in our dispersions using Monte Carlo techniques. 
For each galaxy spectrum we broaden a stellar template to mimic the measured
dispersion, scale the template to the observed cts/pix in the region around
5,200~\AA, and add the sky at the observed level.  We sample this
artificial galaxy spectrum 100 times, introducing the Poisson noise
and sky subtracting, and then measure the dispersion.
We use the RMS of the measured
dispersions around the input value as a measure of the velocity
dispersion uncertainty (see Table \ref{galtab}).  
Not surprisingly, the simulations indicate that averaging the
dispersions measured by cross correlating
against multiple templates does not improve the accuracy.  However,
these simulations consider only Poisson noise effects;
we average the multiple dispersion measurements from each galaxy spectrum
to reduce template mismatch systematics.

We have multiple spectra for 21 galaxies.  The distribution of dispersion
differences scaled by the uncertainties for the 28 comparison pairs
has an RMS of 0.97, indicating that the Monte Carlo uncertainties are
a reasonably good estimate of our true dispersion uncertainties.
The mean scaled difference
($\left<(\sigma_{Mod}-\sigma_{Deca})/\epsilon_\sigma\right>$
where $\epsilon_\sigma$ is the uncertainty in the difference)
between Modspec and Decaspec measurements for 23 comparison pairs
has a mean of $-0.11\pm0.19$ with an RMS scatter about this mean of 0.89;
the variance weighted, mean logarithmic difference between
Modspec and Decaspec dispersions is $-0.001\pm0.013$. 
Because there is no evidence for a systematic
difference between dispersions measured with the Decaspec and the Modspec, 
we apply no correction.
We note that the Modspec was used in measuring dispersions for 10 of
the 16 galaxies in group A and 5 of the 8 galaxies in group B; so any systematic
offset between Decaspec and Modspec measurements would add to the FP scatter
within each group, but would not bias the estimate of the distance difference.

Table \ref{galtab} contains a list of the galaxies and their properties;
for the galaxies with multiple measurements, we use the variance weighted
average dispersion.  The columns of Table \ref{galtab} contain the galaxy
tag (first letter corresponds to group membership), the coordinates,
the redshift $cz$, the velocity dispersion $\sigma$ and uncertainty
$\epsilon_\sigma$, the $R$ band apparent magnitude $M_R$, the mean
surface brightness $\left<\mu_e\right>$ within the half light radius,
the half light radius $\theta_e$, the $Mg_b$ equivalent width, and
the bulge luminosity fraction $F_B$.

For 6 galaxies in J8 we compare our measured dispersions with those of
Saglia {\it et al.} (1997).  There is evidence of an offset;
the distribution of $(\sigma_{EFAR}-\sigma_{obs})/\epsilon_\sigma$
has a mean of $-0.80\pm0.31$ (where $\epsilon_\sigma$ is the uncertainty in
the dispersion difference). 
The average EFAR velocity dispersion is $15\pm7$~km/s
lower than ours.  This offset does not affect our estimates of the relative
distances to the two subclusters, but would complicate efforts to bring our
distances onto the EFAR system.

We place constraints on possible differences in the stellar populations of the
two subclusters by using IRAF scripts to measure $Mg_b$ linestrengths in
each galaxy spectrum.  The rest wavelengths given in Burstein et al. 
(1984) for the feature and continuum bands are corrected using the redshifts in
Table \ref{galtab}.  We broaden our spectra to the nominal
8.6~\AA\ resolution of the Lick system (\cite{wor96b}) before making measurements.
We tranform to the Lick system using an expression of the form
\begin{equation}
Mg_b^{LICK}= a f(\sigma) Mg_b^{obs} + c_{ap},
\end{equation}
where $a$ is a scale factor, $f(\sigma)$ is a velocity dispersion correction
we determine by broadening our Lick standards,
and $c_{ap}$ is an aperture correction of the same form as in Eq. \ref{dispcor}
but with a coefficient of 0.050 instead of 0.038.
We determine $a$ separately for each run; observations of Lick standards indicate
$a=1.060\pm0.016$ for the Nov '96 and $a=1.086\pm0.028$ for the Dec '96 data.
We use overlapping galaxy observations to measure $a=1.066\pm0.034$
for the Nov '95 data.  The aperture correction is
similar to that used by J\o rgensen (1997), but generalized to the form used for
the velocity dispersions above (\cite{bag96}).

We test the accuracy of our $Mg_b$ measurements using multiple galaxy observations.
Twenty three comparison sets from 18 different galaxies yield
an RMS of 0.051 in $\Delta\log{Mg_b}$. 
Thus, we estimate the uncertainty of a single observation
is approximately 9\%.  The linestrengths listed in Table \ref{galtab} are
averages of multiple observations where appropriate.  
Finally, we compare our linestrengths to preliminary EFAR $Mg_b$ linestrengths
in 6 J8 cluster galaxies; the EFAR values are lower:
$\Delta\log{Mg_b}=-0.0733\pm0.014$ with an RMS of 0.039.  The
linestrengths in Table \ref{galtab} are corrected to the EFAR system.

\section{Fundamental Plane Analysis}

In this section we, determine the coefficients and zeropoint
of the FP within the two clusters ($\S$3.1),
discuss the evidence for stellar population variations ($\S$3.2),
and then interpret the observed FP zeropoint offset ($\S$3.3).

\begin{figure}
\plotfiddle{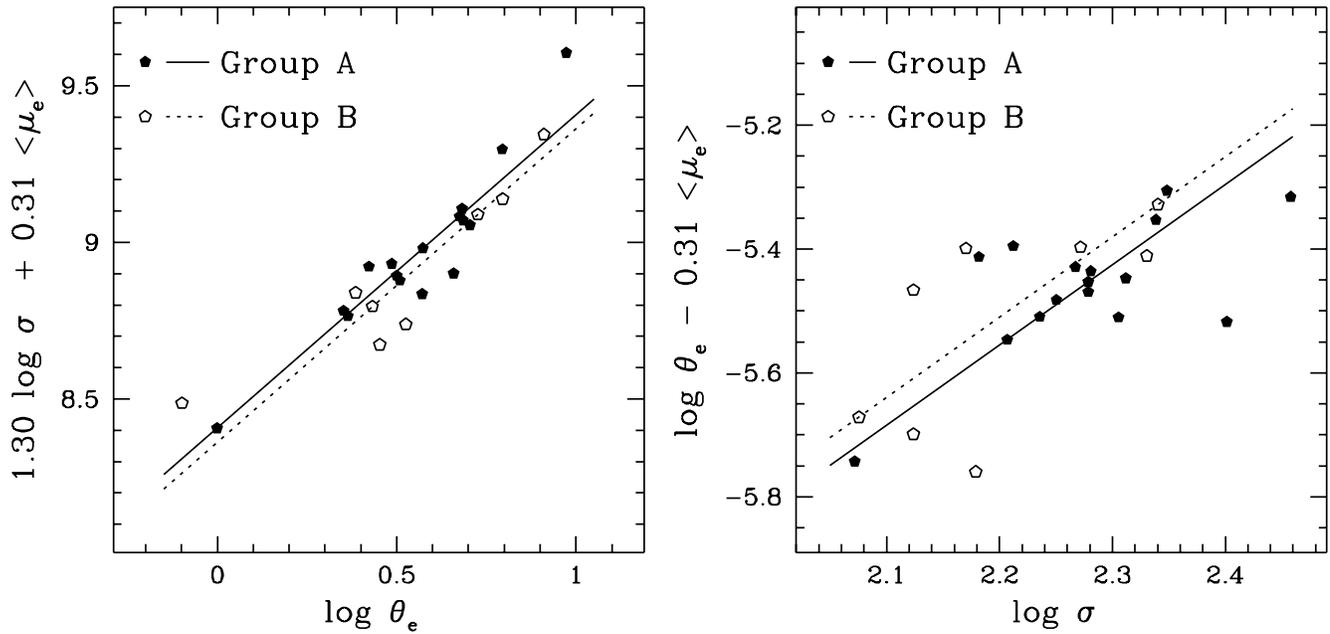}{3in}{0}{95}{95}{-300}{-400}
\caption{Two projections of the FP.  Both plots contain group A (solid) and B (hollow)
galaxies.  The lines define the best fit group A (solid) and B (dashed) FP; offsets
in the fits reflect subcluster distance differences.
The best fit FP (with weighting to account for
incompleteness-- see text) is of the form
$\log{\theta_e}=1.30\log{\sigma}+0.31\left<\mu_e\right>+\gamma_{cl}$ where
$\gamma_{cl}$ is the distance dependent cluster zeropoint.  The RMS scatter around the
best fit in $\log{\theta_e}$ is 0.09, corresponding to a 23\% distance uncertainty
per galaxy.
\label{figFP}}
\end{figure}

\placefigure{figFP}
\subsection{Determining FP Coefficients}

Even if the assumption that elliptical and S0 galaxies within both
clusters are similarly distributed within the FP is valid, the
FP coefficients and cluster distances can be biased by differences in
selection (e.g. \cite{lyn88,bag96}). 
Our galaxy sample is drawn from a list of known elliptical
and S0 cluster members sorted by central aperture magnitude
(aperture is $2''\times2''$ square); the parameters of our final sample
indicate that central aperture magnitude is correlated with
$\log{\theta_e}-0.3\left<\mu_e\right>$, a quantity similar to the
combination which appears in the FP.  Thus, our selection is
similar to selection by isophotal diameter $D_n$ (\cite{dre87}).
We seek to minimally bias the coefficients by (1) combining the samples
from both clusters into a single FP fit and (2) weighting galaxies within
each cluster by the sample completeness.  Finally, we examine the variation of
the zeropoint offset between the two clusters as a function of variations
in the FP coefficients ($\S3.3$).

Our galaxy sample extends over 1.5~mag in central aperture magnitude.  We
divide this range into four bins and calculate the completeness within each 
separately for the two clusters.
The completeness in each bin is simply the number of galaxies within our final
FP sample divided by the estimated total population of elliptical and S0 cluster
members.  We estimate the total cluster population within each bin to be 
$N_{cl}=N_{tot}*f_{cl}$, where $N_{tot}$ is the total
number of elliptical and S0 galaxies identified in the $R$ band
survey of the cluster, and $f_{cl}$ is the fraction of
all elliptical and S0 galaxies with redshifts that lie within the
velocity range of cluster $cl$.

Weighting by the inverse of the completeness appears to be ill advised
for the faintest galaxies within each cluster. 
For these two galaxies the completeness is small enough
(weight is 13.2 (17) for the faintest in group A (B)), that weighting
by the incompleteness would lead to them dominating the fit.
We exclude the faintest galaxy from each sample when determining
the FP coefficients,
but include them when calculating the best estimate of the cluster zeropoint.
Finally, we normalize the weights within each cluster so that the
ratio of weighting
for cluster A:B is 15:7, the ratio of the number of galaxies in each.

We fit the galactic properties to a FP of the form
$\log{\theta_e} = \alpha\log{\sigma}+\beta\left<\mu_e\right> + \gamma$,
where $\alpha$ and $\beta$ are the (cluster invariant) coefficients of
the plane and $\gamma$ is a distance dependent zeropoint
(\cite{fab87,djo87,jor93}, JFK96). We combine
galaxies from both groups in a single fit by using deviations around
median values of distance dependent quantities (\cite{bag96}). 
The corrected quantities $\left<\mu_e\right>$ and $\log{\sigma}$ are
distance independent, so we transform $\log{\theta_e}$ alone; 
specifically, we remove an estimate of
the zeropoint within each subcluster by expressing $\log{\theta_e}$ as the
variation around the $\gamma_{fit}$ which minimizes the cluster scatter in
$\gamma$
\begin{equation}
{\log{\theta_e}\to \log{\theta_e} - \gamma_{fit}}
\end{equation}
Note that the transformation depends on the FP coefficients,
so it must be reapplied during each fit iteration.

We determine the best fit FP coefficients by minimizing
the sum of the absolute value of the weighted, orthogonal deviations
from the plane (JFK96, \cite{bag96}). 
The best fit values are $\alpha=1.30$ and $\beta=0.31$; 
the scatter in $\log{R_e}$ is 0.090 (23\% distance uncertainty).
Our FP has the form
\begin{equation}
\label{FPeq}
{\log{\theta_e} = (1.30\pm0.36)\log{\sigma}+(0.31\pm0.06)\left<\mu_e\right> + \gamma.}
\end{equation}
The uncertainties are determined using a bootstrap resampling
procedure (JFK96, \cite{bag96}) to resample the galaxy list and fit the FP
$10^4$ times; the uncertainty intervals are half the width of the 68\% confidence region.
For comparison, the FP coefficients in the case where all galaxies
are given equal weight are $\alpha=1.27$ and $\beta=0.27$.
Our best fit FP coefficients are statistically consistent with the
values $\alpha=1.24\pm0.07$ and $\beta=0.33\pm0.01$ determined in a sample of
207 galaxies with Gunn $r$ photometry (JFK96); as discussed
by JFK96, their coefficients are consistent with previous
studies (e.g. \cite{fab87,djo87,ben92,guz93,sag93a,jor93}).
Fig. \ref{figFP} contains plots of the group A (filled) and B (hollow)
galaxies along two projections of the FP; the lines in each figure represent
the FP for groups A (solid) and B (dashed).

\begin{figure}
\plotfiddle{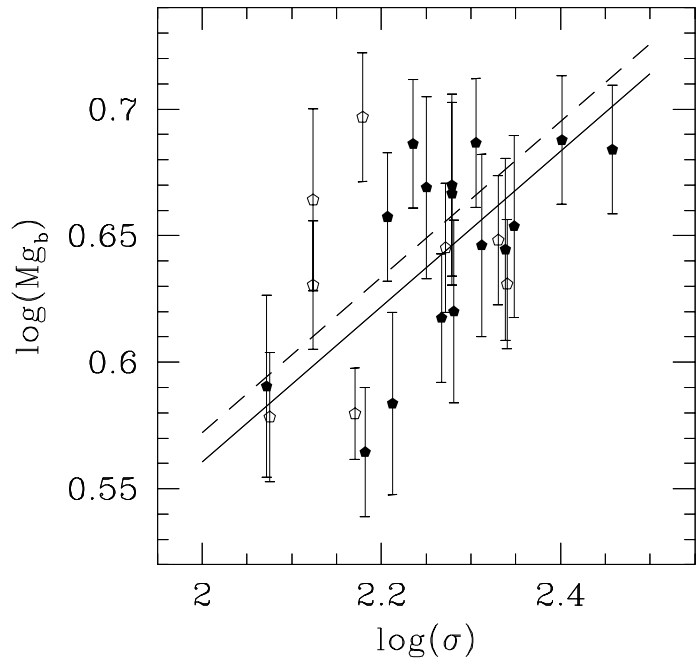}{4in}{0}{100}{100}{-200}{-400}
\caption{Plot of $\log{(Mg_b)}$ vs $\log{(\sigma)}$ for the galaxies in groups
A (solid) and B (hollow).  The lines mark the best fit relations
for groups A (solid) and B (dashed) constrained to have the slope found using
a much larger galaxy sample (J\o rgensen 1997).
The apparent zeropoint offset between the two samples is statistically insignificant
($0.012\pm0.019$; see text), providing no evidence for significant stellar
population differences between the subclusters.
\label{mgfig}}
\end{figure}

\placefigure{mgfig}
\subsection{Constraints on the Stellar Populations}
We compare the distribution of $Mg_b$ equivalent widths within
the two subclusters to place constraints on stellar population variations which
could introduce offsets in the cluster FP zeropoints,
biasing the relative distances to the two clusters (e.g. JFK96).
Fig. \ref{mgfig} contains a plot of $\log{(Mg_b)}$ versus $\log{\sigma}$
for the galaxies in groups A (solid points) and B (hollow points).
The lines represent best fit models for groups A (solid) and B (dashed).

Rather than allowing both the slope and zeropoint to vary, we constrain
the slope to be that found by J\o rgensen (1997) for a sample of
$\sim300$ galaxies.  Specifically, the combined correlations
of $Mg_2$ with $Mg_b$ and $Mg_2$ with $\sigma$ imply
\begin{equation}
\log{Mg_b} = 0.307 \log\sigma - b,
\end{equation}
where the zeropoint for the large galaxy sample is $b=0.034$ (\cite{jor97}). 
By minimizing orthogonal deviations from the line, we find
$b=0.054\pm0.016$ for sample A and $b=0.042\pm0.022$ for sample B.
The zeropoint of each cluster is statistically consistent with
the zeropoint from the J\o rgensen sample.  We can measure
the offset between the two clusters somewhat more accurately;
the zeropoint difference between the
two clusters is $\Delta b=b_A-b_B=0.012\pm0.019$,
statistically consistent with no offset.

Trends in $Mg_2$ have been noted as a function of
radius from the center of the Coma cluster (\cite{guz92}) and
as a function of cluster velocity dispersion for a sample
of 11 clusters (JFK96; \cite{jor97}).
Another study indicates that elliptical galaxies in low
velocity dispersion environments have both higher metallicities
{\it and} younger stellar populations than galaxies in higher
velocity dispersion environments (\cite{ros94}). 
The velocity dispersion difference between clusters A
($\sigma=658$~km/s) and B ($\sigma=415$~km/s) should
introduce an offset of $\Delta b \sim -0.016$ into the $Mg_b$--$\sigma$
correlation.  This expectation is inconsistent with our observations
at the 1.5$\sigma$ level; we would require additional galaxy linestrengths
to resolve an offset of that magnitude.

\subsection{FP Zeropoint Differences}

We determine the zeropoint of each group by using the median $\gamma$;
the 16 galaxies in group A yield $\gamma_A=-8.4083$ with an RMS around this
value of $0.0879$, and the 8 galaxies in group B yield $\gamma_B=-8.3632$
with an RMS of $0.1149$.  The homogeneous Malmquist bias has a
marginal effect on the {\it relative} distances to these two clusters
(\cite{lyn88}); we calculate $\delta\gamma_A=-0.0033$ and
$\delta\gamma_B=-0.0114$, resulting in an increase in $\Delta\gamma$ of
$\sim0.0081$.  Thus, the zeropoint difference between the two clusters is
\begin{equation}
\label{gamoff}
\Delta\gamma=\gamma_A-\gamma_B=-0.037\pm0.046,
\end{equation}
where we have assumed the uncertainty in the zeropoint
is the scatter about that value
divided by $\sqrt{N}$, where $N$ is the number of galaxies (note that this
is a conservative estimate of the true uncertainties because the scatter is
sensitive to outliers).
As discussed in $\S2.2$, if we assume groups A and B are at the same distance
rather than at distances proportional to their mean redshifts, the cosmological 
dimming corrections change so that $\Delta\gamma$ increases by $\sim0.004$.

\begin{figure}
\plotfiddle{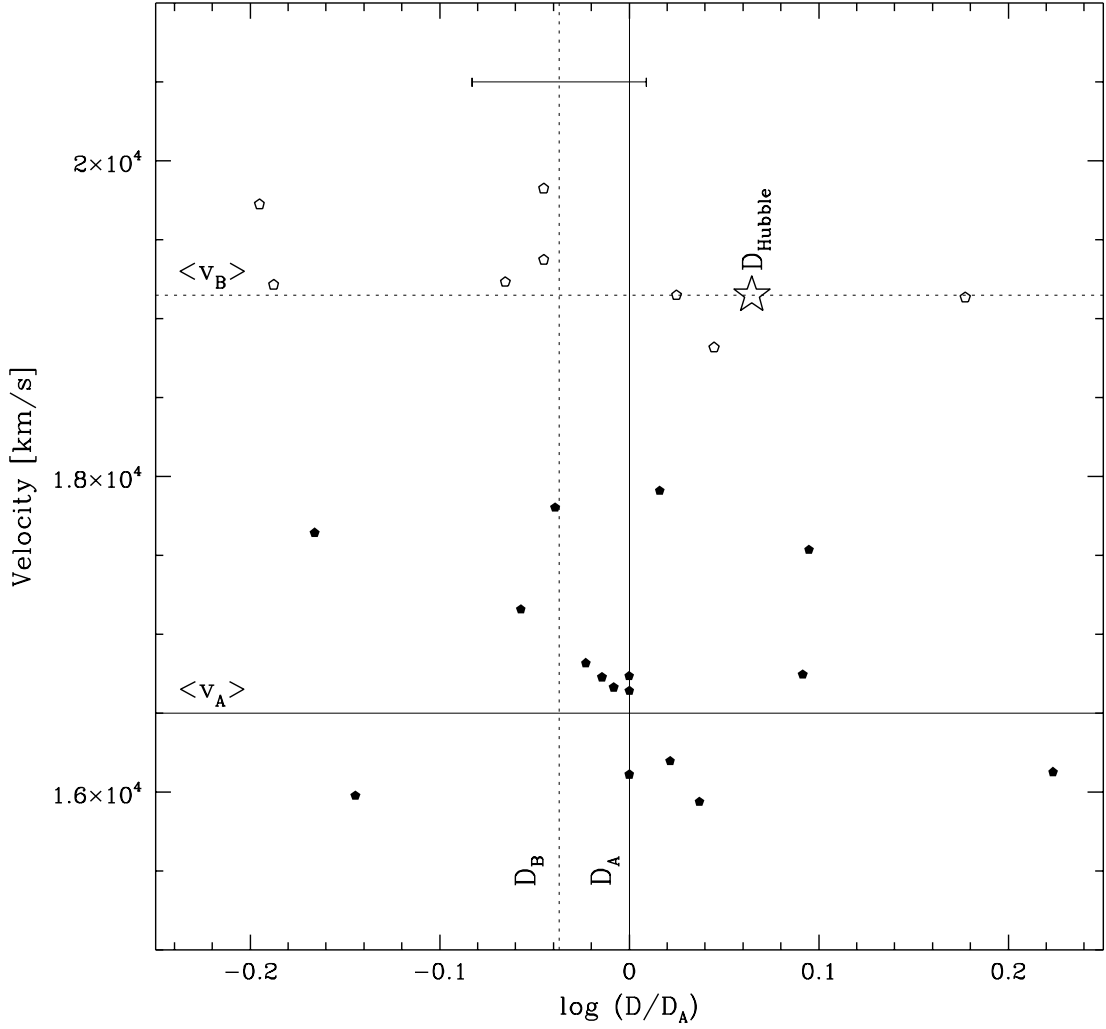}{5in}{0}{80}{80}{-250}{-150}
\caption{Plot of fundamental plane relative distances versus redshift for
galaxies in groups A (solid) and B (hollow).  Galaxies in groups A and B are
clearly offset in velocity (see Fig. \ref{figvel}), and also appear
to be offset in distance.
Vertical lines mark the (homogeneous Malmquist bias corrected) median distances
to groups A (solid) and B (dashed). 
Horizontal lines mark the mean velocities of groups A
(solid) and B (dashed).  The large star marked with ${\rm D_{Hubble}}$ lies at
the expected position of group B if its trajectory relative to group A
were pure Hubble flow.  The median distances to groups A and B indicate that
$\log{\left({\rm D_B/D_A}\right)}=-0.037\pm0.046$;
the $1\sigma$ distance uncertainty
is indicated by the error bar on the vertical, dashed line.
The sample of 24 galaxies rules out the
Hubble flow hypothesis, ${\rm D_B=D_{Hubble}}$ and
$\log{\left({\rm D_B/D_A}\right)}=0.065$, with 99\% confidence (see $\S3.3$). 
\label{figdistance}}
\end{figure}

\placefigure{figdistance}

In general, zeropoint offsets can be caused by
stellar population or metallicity induced $M/L_R$ differences and/or
distance differences.  Because there is no compelling evidence
for stellar population differences, we use the observed
zeropoint offset and uncertainty as relative distance constraints. 
The quantity $\gamma$ (in Eqn. \ref{FPeq})
can be written
$\gamma_x=\log{\left(\Gamma/D_x\right)}$ where $D_x$ is the distance
of galaxy $x$ and $\Gamma$ follows from the standard form implied by the FP
\begin{equation}
{R_e=\Gamma\sigma^{\alpha}\left<I_e\right>^{-2.5\beta}.}
\end{equation}
Assuming the intrinsic FP zeropoint $\log\Gamma$
is the same for all galaxies, the relative distances to two galaxies
$x$ and $y$ is
\begin{equation}
{\log{\left(D_x\over D_y\right)} = \gamma_y - \gamma_x.}
\end{equation}
\noindent
Fig. \ref{figdistance}
contains a plot of the distances to the
galaxies in groups A (solid) and B (hollow) relative to
the median distance to group A versus
the galaxy redshift.  Also marked with vertical
lines are the (homogeneous Malmquist bias corrected) median distances
to groups A ($D_A$; solid line) and B 
($\log\left(D_B/D_A\right)=-0.037$; dashed line); the horizontal lines mark the
mean velocities for groups A (solid) and B (dashed).  The error bars on
the dashed, vertical line represents the relative distance uncertainty.
The large star marks the position of cluster B if its distance relative to
cluster A were reflective of pure
Hubble flow ($\Delta D\propto\Delta\left<v\right>$);
the Hubble flow hypothesis implies $\log\left(D_B/D_A\right)=0.065$. 

The zeropoint difference is consistent with zero, and it is sufficient
to rule out Hubble flow at $2.2\sigma$;
therefore, the FP measurements
suggest that cluster B is closer to us than cluster A
(in $\S5.1$ we return to this issue in more detail).
Naturally, the zeropoint offset depends on the coefficients of the FP,
which are only constrained to within $\sim$20\% of the best fit values.
Using the distribution of $\gamma_A-\gamma_B$ from the $10^4$
bootstrap resampling fits, we find that in 99.05\% of the simulations
the zeropoint offset between the two clusters is less than the Hubble
flow offset, rejecting the Hubble flow hypothesis.  This probability
corresponds to a $+2.35\sigma$ deviation in a Gaussian, similar to the
2.2$\sigma$ estimate above which we derived from the scatter about the median
distance in each cluster.  The bootstrap probability is superior because
it includes variations in the FP coefficients (and is less sensitive to
outliers than the scatter estimator).

\section{Radial Infall Model as a Mass Constraint}

The radial infall model has been applied
to many close galaxy cluster pairs to determine whether or not
the pairs will merge (e.g. \cite{bee82,bee91,bee92,col95,sco95}). 
In the following
we use the measured line of sight separation and this
model to estimate the total binding mass required to
explain the observed infall velocity.  We then compare this mass to
virial estimates of the cluster binding mass.

\subsection{Radial Infall Model}
For two isolated, bound objects within an expanding universe, the dynamics
can be parametrized in the standard way
\begin{eqnarray}
t=B(\eta-\sin\eta)\\
l=\left(GM_GB^2\right)^{1/3}(1-\cos\eta)\nonumber\\
v= \left({GM_G\over B}\right) ^{1/3} {\sin\eta\over (1-\cos\eta)}\nonumber
\end{eqnarray}
where $\eta$ is a development angle, $l$ is the separation, $v$ is the relative velocity,
$t$ is the age of the universe, $B$ is an undetermined coefficient,
and $M_G$ is the sum of the individual masses (e.g.\ \cite{pee93}).
When we observe two clusters on a merger trajectory at the present epoch,
we measure the
projected separation of the clusters
$\Delta l_\perp$ and a line of sight velocity difference $\Delta v_{LOS}$.
Modelling the collision as a radial infall, we parametrize the
merger in terms of the angle $\phi$ between the merger trajectory
and the line of sight ($\Delta v_{LOS}=\Delta v\cos\phi$ and
$\Delta l_\perp=\Delta l\sin\phi$).

With four free parameters in this model ($\eta$, $B$, $\phi$ and $M_G$)
and three observables (the age of the universe $t_H$, the 
projected separation $\Delta l_\perp$ and the line of sight velocity difference
$\Delta v_{LOS}$), the total cluster mass and
age of the universe are a function of the angle $\phi$ and
the present epoch value of the development angle $\eta_H$.
\begin{eqnarray}
\label{massfunc}
M_G= 2.3\times10^{14}M_\odot \left((1+\cos\eta_H)\cos^2\phi\sin\phi\right)^{-1}\nonumber\\
\left({\Delta v_{LOS}\over 1000\ {\rm km/s}}\right)^2 \left({\Delta l_\perp\over 1\ {\rm Mpc}}\right)\nonumber\\
\\
t_H=9.8\times10^8\ {\rm yrs}\ \left({(\eta_H-\sin\eta_H)\sin\eta_H\over(1-\cos\eta_H)^2} \cot\phi\right)\nonumber \\
\left({\Delta l_\perp\over 1\ {\rm Mpc}}\right) \left({1000\ {\rm km/s}\over \Delta v_{LOS}}\right)\nonumber
\end{eqnarray}
\noindent Using an approximate age of the universe to determine $\eta_H$,
we arrive
at the total cluster mass function $M_G(\phi)$.
Fortunately the dependence of $M_G$ on $t_H$ is rather
weak (see Fig. \ref{figmerge}).
If the line of sight separation between the subclusters $\Delta l_{LOS}$
is measured, $\phi$ is determined, yielding the total system mass
(which we term the cluster merger mass $M_G$).
$M_G$ can then be compared to other mass estimates, which
typically probe the cores of clusters.

Departures from radial infall and the physical
extent of the clusters can bias the merger mass.  
Generally speaking, the large mass and rarity of
galaxy clusters make them good candidates for this analysis.
We plan to study the
accuracy of merger masses using numerical simulations of cluster evolution
within ``realistic'' environments.

Note that there are
scenarios in which the merger mass would be substantially biased; it is
possible to recognize these scenarios observationally.  
In the first scenario, the two merging
clusters are at small separation 
($\le1$~Mpc); interactions among the cluster components, significant overlap
of their mass profiles, and small departures from pure radial infall
combine to bias the inferred merger velocity and the implied merger mass.
Fortunately, the signatures of cluster mergers have been extensively
studied (e.g. \cite{gel82,dre88,jon92,moh93,pea94,moh95,buo95}), and typically
there are merger clues in the galaxy and gas distributions.

In the second scenario, a third object of comparable mass to
the two merging clusters is close enough to invalidate the two body analysis.
In this case, the observed distribution of nearby galaxy clusters
should be enough to determine whether or not the radial infall model is
appropriate.

\placefigure{figmerge}
\subsection{Merger Mass in A2626}

To calculate $M_G$ we require (see Eqn. \ref{massfunc}) the observed
rest frame velocity difference and the projected separation.
The rest frame velocity difference is $\Delta v=2,486\pm112$~km/s.
The projected separation between the centers of mass
is assumed to be the distance between the bright, central elliptical in group A
(its position is consistent with the peak in the X--ray emission)
and the centroid of the 30 group B members with measured redshifts
(using the centroid of group A members slightly 
increases the projected separation and the merger mass).
At a cluster distance of $l=175h^{-1}$~Mpc, the projected separation between
the two groups is $\Delta l_\perp=(0.707\pm0.052)h^{-1}$~Mpc.
The uncertainty reflects the statistical
uncertainties associated with centroiding group B.
The fractional uncertainty in the merger mass $M_G$ due to
$\Delta v_{LOS}$ and $\Delta l_\perp$ is 12\%.

Fig. \ref{figmerge}
displays the ratio of the cluster merger mass $M_G$ to the sum of the virial
masses of groups A and B ($M_A+M_B=9.1\times10^{14}h^{-1}M_\odot$;
see Table \ref{grouptab} and MGW96)
versus the line of sight separation between the subclusters,
$\Delta l_{LOS} = \Delta l_\perp / \tan{\phi}$. 
The dotted line corresponds to $t_H=18$~Gyr and the solid line corresponds
to $t_H=13$~Gyr.
The FP distances to galaxies within groups A and B yield a ratio of the
group B distance to
the group A distance ($l_B/l_A$) which can be used to
calculate $\Delta l_{LOS}$.
Specifically, we estimate the line of sight separation between the subclusters
as $\Delta l_{LOS}\sim 175 h^{-1} (1-l_B/l_A)$~Mpc.
The sample of 24 FP distances to groups A and B
yields $\Delta l_{LOS}= (14\pm18)h^{-1}$~Mpc.

Because distance uncertainties in this case are large, our method does not yield
tight mass constraints in Abell~2626;  however, 
if groups A and B are merging (99\% confidence from FP distances and
93\% confidence from galaxy magnitude
distributions), it is clear that the virial masses of
groups A and B underestimate the
total system binding mass.  Specifically, the sum of the masses of groups
A and B must be at least $1.5\times10^{15}M_\odot$, compared to the virial
sum of $9.1^{+3.1}_{-2.7}\times10^{14}M_\odot$ (MGW96), a factor of 1.65 higher.
The virial mass range corresponds to 90\% statistical confidence limits,
where it is assumed the mass uncertainties are dominated by the
velocity dispersion uncertainties (\cite{hei85}).
The virial sum is inconsistent with the {\it minimum} merger mass at
the $\sim3\sigma$ level.

\section{Discussion}

The study of double clusters with the fundamental plane
holds promise for significant progress in two areas.
First, the study of appropriately isolated nearby double clusters should
lead to new constraints on the total cluster binding mass.
These constraints will not only provide
an independent test of the virial, hydrostatic, and weak lensing mass
estimators, but will also provide information about the mass (and typical
$M/L$) outside the virialized cluster region.  Second,
the study of more distant double clusters
(at $z\ge0.1$) provides a means of directly probing for environmentally
driven biases in FP distance estimates.

\subsection{Fundamental Plane Analysis}

Here we apply this approach to the double cluster
A2626 at $cz\sim17,500$~km/s.  Because of the cluster distance, it is critical
to make seeing corrections while extracting $\theta_e$ and
$\left<\mu_e\right>$.  Our method is similar to the
one described by Saglia {\it et al.} (1993b), but
includes a seeing--correction uncertainty which makes the photometric
parameters  less sensitive to the in/exclusion of central points in the profile.
We also note that the cosmological dimming
correction to the surface brightness has a peculiar velocity dependence which
may be important in cases where many FP distance estimates are combined to
produce a single, more accurate cluster distance ($\S2.2$).  

Through numerous cross checks we demonstrate that
our estimates of the combination
$\log(\theta_e)-\beta \left<\mu_e\right>$ which appears in the FP are
sufficiently accurate that their errors make no significant contribution
to the FP scatter;  using a Monte Carlo approach we
estimate the velocity dispersion
uncertainties ($\S2.3$).  These parameters for the 8 galaxies in group
B and the 16 in group A are listed in Table \ref{galtab}.

We combine galaxies from both clusters in a single fit, weighting to account
for incompleteness ($\S3.1$).  Our best fit FP coefficients
($\alpha=1.30\pm0.36$ and $\beta=0.31\pm0.06$) are
statistically consistent with measurements in other
clusters (see Fig. \ref{figFP}).  The RMS scatter about this plane in
$\Delta\log{R_e}$ is 0.09, corresponding to 23\% distance uncertainties
per galaxy.

We examine the distribution of $Mg_b$ equivalent widths with the galaxy
spectra from both subclusters ($\S3.2$). 
We find that there is no evidence for significant
differences in the $Mg_b$--$\sigma$ relation (see Fig. \ref{mgfig}). 
Our result does not contradict the recently noted correlation between
cluster environment and linestrength (JFK96, \cite{jor97}); 
the expected offset in $Mg_b$ (given the velocity dispersion differences
between groups A and B) is too small to detect with our data.

The difference in the fundamental plane zeropoints for the two clusters is
$\gamma_A-\gamma_B=-0.037\pm0.046$, consistent with no offset
($\S3.3$).  Under the
assumption that this zeropoint difference is indicative of distance
differences, $\log{(D_B/D_A)}=-0.037\pm0.046$, where the uncertainty
follows from scatter around the median distance in each cluster with the
best fit FP coefficients.
We use the $10^4$ bootstrap resampling simulations to measure the
variation of the distance offset as FP coefficients vary.
The relative distance constraint is robust enough
to reject the Hubble flow hypothesis ($\log{(D_B/D_A)}=0.065$)
in 99\% of the simulations. Under the Hubble flow hypothesis,
the two clusters are not interacting gravitationally,
and the velocity difference reflects pure Hubble flow.  
An analysis of the $R$ band magnitude distributions of confirmed
members of both clusters rules out Hubble flow with 93\% confidence (MGW96).

If the subclusters were on an outgoing trajectory (bound or unbound),
then the line of sight distance difference would have to be greater
than their current recession velocity times the age of the universe
($\Delta l_{LOS}>v_{LOS}t_H\sim34$~Mpc) which is
greater than the Hubble flow value: $\log{(D_B/D_A)}\ge0.0645$; 
thus, models where groups A and B are on outgoing trajectories are
ruled out with higher confidence than
pure Hubble flow.  Therefore, our relative distances indicate that the
subclusters are bound and infalling with 99\% confidence
(see Fig. \ref{figdistance}).

\subsection{Radial Infall Model}

\begin{figure}
\plotfiddle{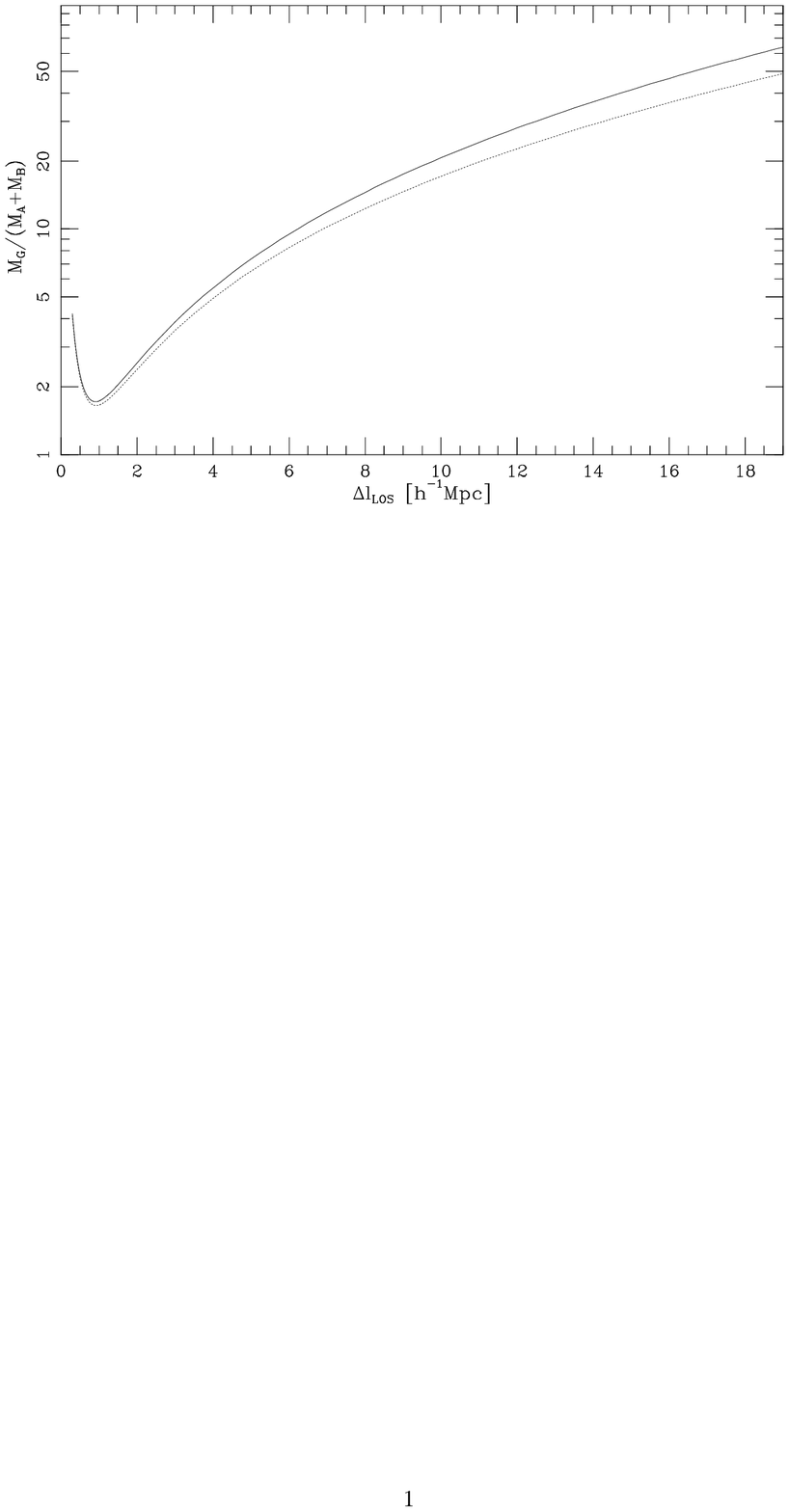}{5in}{0}{130}{130}{-400}{-600}
\caption{
The ratio of the cluster merger mass $M_G$ to the
sum of the group A and B virial masses $M_A+M_B=9.1\times10^{14}h^{-1} M_\odot$
as a function of the line of sight separation $\Delta l_{LOS}$ of the
two groups.  The two solutions correspond to different assumptions
regarding the age of the universe $t_H$: $t_H=13$~Gyr (solid) and $t_H=18$~Gyr (dotted). 
The statistical uncertainty in the merger mass $M_G$ is 12\%.  If groups
A and B are merging, as suggested by the FP distances and the galaxy magnitude
distributions, the merger mass is at least 1.65 times greater than
the sum of the virial masses.  The FP distances indicate that
$\Delta l_{LOS}=14\pm18\ h^{-1}$~Mpc.
\label{figmerge}}
\end{figure}

Assuming bound and infalling subclusters,
we model these kinematics with a radial infall model
(\cite{bee82}); this two body model is appropriate for
separations large compared to the scale
of the virialized region of the cluster and in cases where there are no
other massive clusters in the neighborhood.  Given the projected
separation, line of sight velocity difference, and line of sight separation,
the radial infall model provides an estimate of the total cluster binding mass
which depends weakly on the adopted age of the universe.
From Fig. \ref{figmerge}, the total binding mass is at least 1.65 times
higher than the virial sum, a difference significant at $\sim3\sigma$.
The total binding mass could be much larger than this minimum.

The differences in these two mass estimates could indicate
(1) deviations from radial infall, (2) an error in the virial estimate,
or (3) significant mass beyond the virialized region. 
The X--ray image of A2626 provides no indication of
interactions between the two subclusters, and there is no third subcluster,
so the radial infall model should be appropriate in A2626.
If the virial estimators are biased low,
then the radial infall model indicates that $M/L_R\sim1000h$
(see Table \ref{grouptab}), which would be a surprisingly high
value (e.g. Ramella et al. 1989).  These arguments suggest that
the minimum merger mass is larger than the virial mass
because it is sensitive to mass in the cluster infall region.

Taking the merger mass as a lower limit on the mass within the cluster
infall region, we can address whether there is evidence for variation 
of the mass to light ratio between the cluster core and infall region.
The galaxy light is calculated within a region which
only extends to a projected radius of 1.5$h^{-1}$~Mpc; presumably
both the galaxy and mass distributions extend beyond this region. 
We have no CCD photometry over the larger region
to directly measure the total cluster light.  Assuming that the projected
galaxy distribution falls off as $\sim R^{-1.5}$ (e.g.\ \cite{moh96a})
with a similar luminosity function to the galaxies in the central region,
the total cluster light grows as $\sim\sqrt{R}$ for $R$ larger than the core
radius.  Under these assumptions, within a diameter
of $\sim4h^{-1}$~Mpc there is
enough light to yield a mass--to--light ratio of
$M/L_R\sim600h$ throughout the cluster even with the
higher mass estimate from the radial infall model.  Of course, the
binding mass could be many times larger than the minimum value,
so the mass--to--light ratio could increase significantly outside
the virialized cluster core.  A study of the cluster light extending
to larger radius and a larger sample of FP distances would yield more
concrete information regarding any variation
in the cluster mass--to--light ratio in A2626.

FP studies of other double clusters, coupled with (1) a more detailed
accounting of the radial distribution of cluster light and (2) larger redshift
samples to enable a more accurate estimate of the virial mass, should provide
interesting constraints on the distribution of cluster mass and possible
differences in the efficiency of galaxy formation beyond the virialized region.

\acknowledgements

We thank Roberto Saglia and Glenn Baggley for their comments on an
earlier version of this manuscript and for their generosity in allowing
us to compare our measurements to their unpublished results.
We thank Guy Worthey for stimulating discussions,
Paul Schechter for obtaining improved 2.4~m images of
several galaxies in our sample, and Emilio Falco for attempting additional
imaging. This research was funded in part by NAGW--2367 and NAG5--3401.

\onecolumn

\begin{deluxetable}{lccccc}
\tablewidth{0pt}
\tablecaption{Subcluster Properties}
\tablehead{
\colhead{ID}		& 
\colhead{N}	& 
\colhead{$\bar v$} 	&
\colhead{$\sigma$}	&
\colhead{$M_{vir}$[$10^{14}M_\odot$]} &
\colhead{$M_{vir}/L_R$}}
\startdata
A 	&67	& 16,533$\pm$141 &  658$^{+111}_{-81}$  & 6.6$^{+2.4}_{-1.5}$ & 630\nl
B 	&30	& 19,164$\pm$138 &  415$^{+117}_{-72}$  & 2.3$^{+1.4}_{-0.7}$ &570\nl
\enddata
\tablenotetext{}{Intervals are statistical 90\% confidence limits}
\label{grouptab}
\end{deluxetable}

\begin{deluxetable}{lcccccccccc}
\tablewidth{0pt}
\tablecaption{Galaxy Properties}
\tablehead{
\colhead{Tag}			&
\colhead{RA(1950)}		&
\colhead{Decl}			&
\colhead{$cz$}			&
\colhead{$\sigma$}		&
\colhead{$\epsilon_\sigma$}	&
\colhead{$M_R$}			&
\colhead{$\left< \mu_e\right>$}	&
\colhead{$\theta_e^a$}		&
\colhead{$Mg_b$}		&
\colhead{$F_B^b$}}
\startdata
AA	& 23 34 29.30 & 20 22 22.3 & 17,538 & 287 & 12 & 14.06 & 19.79 & 6.232 & 4.83 & 0.53 \nl
AB	& 23 35 18.21 & 20 31 56.7 & 17,159 & 223 & 22 & 13.93 & 19.47 & 5.477 & 4.50 & 0.34 \nl
AC	& 23 34 22.71 & 20 56 55.6 & 16,819 & 218 & 22 & 14.37 & 19.56 & 4.846 & 4.41 & 0.79 \nl
AD	& 23 33 40.11 & 20 45 31.4 & 17,911 & 190 & 22 & 14.75 & 19.93 & 4.817 & 4.64 & 0.63 \nl
AE	& 23 33 23.09 & 20 42 02.1 & 15,941 & 205 & 28 & 15.03 & 19.22 & 3.063 & 4.42 & 0.55 \nl
AF      & 23 35 33.91 & 21 07 54.7 & 16,728 & 191 & 12 & 14.97 & 19.23 & 3.161 & 4.17 & 0.74 \nl
AG      & 23 32 40.07 & 21 00 01.6 & 15,979 & 163 & 19 & 14.71 & 19.33 & 3.727 & 3.83 & 0.85 \nl
AH	& 23 33 40.68 & 20 42 35.6 & 16,644 & 161 &  6 & 15.02 & 20.15 & 4.732 & 4.54 & 0.50 \nl
AI	& 23 34 32.46 & 20 18 14.5 & 17,804 & 185 &  7 & 14.93 & 19.24 & 3.231 & 4.14 & 0.84 \nl
AJ	& 23 34 35.93 & 20 59 41.0 & 16,665 & 178 & 35 & 15.35 & 18.94 & 2.319 & 4.67 & 0.90 \nl
AK	& 23 33 58.71 & 21 01 12.4 & 16,856 & 202 & 20 & 15.35 & 19.23 & 2.564 & 4.86 & 0.70 \nl
AL	& 23 34 08.75 & 20 49 29.9 & 17,645 & 152 &  3 & 14.61 & 19.67 & 4.554 & 3.67 & 0.39 \nl
AM	& 23 33 49.52 & 20 49 12.1 & 16,129 & 252 & 21 & 14.40 & 21.03 & 9.403 & 4.87 & 0.64 \nl
AO	& 23 33 57.86 & 20 51 11.8 & 16,113 & 190 & 27 & 15.91 & 17.66 & 0.997 & 4.68 & 1.00 \nl
AP	& 23 33 47.81 & 20 49 21.8 & 16,859 & 172 &  8 & 15.07 & 19.70 & 3.743 & 4.86 & 0.30 \nl
AZ	& 23 35 19.60 & 20 30 15.3 & 16,063 & 118 & 11 & 16.02 & 19.74 & 2.136 & 3.89 & 0.70 \nl
BA	& 23 34 07.96 & 20 31 29.6 & 19,729 & 148 &  3 & 14.84 & 19.19 & 3.351 & 3.80 & 0.69 \nl
BB	& 23 33 19.80 & 20 33 12.0 & 19,369 & 219 & 21 & 14.31 & 19.61 & 4.900 & 4.27 & 0.74 \nl
BC	& 23 33 57.29 & 20 35 11.2 & 19,240 & 187 & 17 & 14.36 & 20.06 & 6.243 & 4.42 & 0.69 \nl
BD	& 23 33 08.31 & 20 48 02.8 & 19,150 & 214 & 10 & 14.21 & 20.48 & 8.148 & 4.45 & 1.00 \nl
BE	& 23 33 17.67 & 20 22 54.2 & 19,215 & 133 &  3 & 15.18 & 19.17 & 2.836 & 4.27 & 0.62 \nl
BF	& 23 33 33.61 & 20 49 36.5 & 19,825 & 119 &  3 & 15.89 & 19.77 & 2.705 & 3.79 & 0.38 \nl
BG      & 23 33 17.50 & 20 52 03.1 & 18,818 & 133 &  6 & 16.02 & 19.71 & 2.431 & 4.61 & 0.68 \nl
BH	& 23 33 12.20 & 20 35 20.1 & 19,136 & 151 &  5 & 17.09 & 18.34 & 0.761 & 4.98 & 0.33 \nl
\enddata
\tablenotetext{a}{in arc seconds}
\tablenotetext{b}{bulge luminosity fraction}
\label{galtab}
\end{deluxetable}

\appendix

\section{Effects of Culling the Galaxy Sample}

Four of our 24 galaxies show peculiar morphologies;
here we detail the peculiarities and demonstrate that
eliminating them from the analysis does not significantly
alter our conclusions.
The following are morphological types which we derived by comparing
deep 2.4~m images and the Carnegie Atlas of Galaxies (\cite{san94}):

\begin{itemize}
\item AH: bright nucleus with possible bar; probable SB0
\item AL: possible spiral structure; SB0 or Sa
\item AM: bright nucleus with possible bar; probable SB0
\item AP: appears to have a dust ring; probable RSB0
\end{itemize}

Without these four galaxies, Group A (B) contains 12 (8) galaxies; we
now apply the analysis detailed in $\S3$ to this smaller sample.
The best fit form of the FP is
\begin{equation}
{\log{\theta_e} = (1.39\pm0.65)\log{\sigma}+
(0.32\pm0.06)\left<\mu_e\right> + \gamma,}
\end{equation}
and the scatter around this plane in $\log{R_e}$ is 0.080 (20\% distance
uncertainty).  These coefficients are consistent with the
values from the full sample, but the scatter is somewhat
smaller and the $\alpha$ uncertainty is larger. 
The FP coefficient uncertainties are determined (as before)
through bootstrap resampling and refitting the sample.

The RMS scatter around the median distance estimator in group
A (B) is 0.066 (0.115); the zeropoint difference corrected for 
homogeneous Malmquist bias is
\begin{equation}
\Delta\gamma=-0.026\pm0.045.
\end{equation}
The zeropoint difference and uncertainty (from the scatter)
are somewhat smaller than
for the whole sample (see Eqn. \ref{gamoff}).  This
measurement is inconsistent with the Hubble flow hypothesis at $\sim2.0\sigma$
or 97.7\% confidence.  The bootstrap
refitting produces a distribution of $\Delta\gamma$ which includes the
variation in the FP coefficients;  5000 refits of this smaller sample
rule out the Hubble flow hypothesis with 98.2\% confidence.
In summary, removing 4 galaxies which have morphological peculiarities
slightly reduces the significance with which the Hubble flow model can
be rejected in Abell 2626.

\end{document}